\documentclass[runningheads]{llncs}
\usepackage[T1]{fontenc}
\usepackage{graphicx}
\usepackage{color}
\usepackage{bbm}
\usepackage{multirow}
\usepackage[inline]{enumitem}
\usepackage{subfigure} 
\usepackage{sistyle}
\usepackage[ruled]{algorithm2e}
\usepackage{booktabs}
\usepackage{caption}
\SIthousandsep{,}
\usepackage{makecell}
\usepackage[skip=0pt]{caption}
\usepackage{amsmath}
\usepackage{hyperref}
\usepackage{array} 
\usepackage{graphicx}
\usepackage{xcolor}
\usepackage{acronym}
\usepackage{amsfonts}   
\usepackage{amssymb}    

\newcommand{\heading}[1]{\vspace*{0.5mm}\noindent\textbf{#1.}}

\begin{document}
\title{A Comparative Study of Specialized LLMs as Dense Retrievers} 
%
%
%

\author{Hengran Zhang\inst{1,2,3}\orcidID{0009-0004-1144-1298} \and
Keping Bi\inst{1,2,3}\orcidID{0000-0001-5123-4999} \and
Jiafeng Guo\inst{1,2,3}\orcidID{0000-0002-9509-8674}}

\authorrunning{Hengran Zhang et al.}

\institute{
Key Laboratory of Network Data Science and Technology, Institute of Computing Technology, Chinese Academy of Sciences, Beijing, China \and 
State Key Laboratory of AI Safety, Beijing, China \and
University of Chinese Academy of Science, Beijing, China \\
\email{\{zhanghengran22z, bikeping, guojiafeng\}@ict.ac.cn}}

\maketitle             
\begin{abstract}
While large language models (LLMs) are increasingly deployed as dense retrievers, the impact of their domain-specific specialization on retrieval effectiveness remains underexplored. 
This investigation systematically examines how task-specific adaptations in LLMs influence their retrieval capabilities—an essential step toward developing unified retrievers capable of handling text, code, images, and multimodal content. 
We conduct extensive experiments with eight Qwen2.5 7B LLMs, including base, instruction-tuned, code/math-specialized, long reasoning, and vision-language models across zero-shot retrieval settings and the supervised setting. 
For the zero-shot retrieval settings, we consider text retrieval from the BEIR benchmark and code retrieval from the CoIR benchmark. 
Further, to evaluate supervised performance, all LLMs are fine-tuned on the MS MARCO dataset. 
We find that mathematical specialization and the long reasoning capability cause consistent degradation in three settings, indicating conflicts between mathematical reasoning and semantic matching.  
The vision-language model and code-specialized LLMs demonstrate superior zero-shot performance compared to other LLMs, even surpassing BM25 on the code retrieval task, and maintain comparable performance to base LLMs in supervised settings. 
These findings suggest promising directions for the unified retrieval task leveraging cross-domain and cross-modal fusion.\footnote{\url{https://github.com/Trustworthy-Information-Access/A-Comparative-Study-of-Specialized-LLMs-as-Dense-Retrievers}.}

\end{abstract}

\section{Introduction} 
First-stage retrieval plays a crucial role in information access, serving as the foundation for downstream processes such as re-ranking or augmenting generation. Its primary goal is to retrieve as many relevant documents as possible from a large-scale corpus, which may contain millions or even billions of entries. Traditionally, term-based retrieval methods like BM25 \cite{robertson2009probabilistic} have faced challenges such as vocabulary mismatch due to their reliance on lexical matching between query and document terms. Over time, retrieval techniques have evolved from these classical term-based approaches to neural retrievers like DRMM \cite{guo2016deep}. More recently, retrievers based on pre-trained language models (PLMs) have become prevalent. Known as dense retrieval, PLM-based retrieval typically employs a dual encoder mechanism that encodes queries and documents into embeddings, assessing their relevance using functions like dot product or cosine similarity \cite{karpukhin2020dense}.

The emergence of large language models (LLMs) such as Qwen series~\cite{yang2024qwen2}, GPT series~\cite{gpt-4}, and DeepSeek~\cite{deepseek} has shown compelling language understanding and generation capabilities. 
This makes it appealing to researchers to study how to leverage them for dense retrieval. Recent studies indicate that LLMs also show strong potential in dense retrieval tasks. For instance, Llama fine-tuned with MS MARCO significantly outperforms BERT-based retrievers~\cite{ma2024fine}. 

There are various types of LLMs tuned specifically for different tasks or domains, such as those specialized for code generation~\cite{hui2024qwen2coder}, mathematical reasoning \cite{yang2024qwen2math}, text generation \cite{yang2024qwen2}, multi-modal tasks \cite{qwen2.5vl}, and so on.
Although LLMs have been shown to be powerful as dense retrievers, their specialized capabilities could have a different impact on their retrieval performance if they are used as dense retrievers. For example, do the coding or mathematical reasoning capability negatively impact LLMs' text retrieval potential? Will more competent LLMs, such as long-reasoning and multi-modal models, be better backbones as denser retrievers? Studies on these questions will provide insights into understanding the impact of LLMs' specialty on their capabilities as dense retrievers. This is also beneficial for developing a unified retriever that can handle the retrieval of text, code, image, etc.  
Therefore, in this paper, we present a comprehensive investigation into the relationship between LLMs' specialty and their retrieval performance. 

To ensure methodological rigor and comparative fairness in our experiments, we conducted a comprehensive evaluation of domain-specialized language models using eight Qwen2.5 variants as retrieval backbones, i.e., Qwen-2.5-7B, Qwen-2.5-7B-Instruct, Qwen2.5-7B-Coder, Qwen2.5-7B-Coder-Instruct, Qwen2.5-7B-Math, Qwen2.5-7B-Math-Instruct, Qwen2.5-7B-VL-Instruct and Deepseek-R1-distill-Qwen-7B. 
We employ three settings:  1. zero-shot text retrieval setting (low-resource datasets from the BEIR benchmark \cite{beir}); 2. zero-shot code retrieval setting (CoIR benchmark \cite{CoIR}); 3. supervised training setting (MS MARCO and TREC DL19/20). 
Our systematic investigation yielded the following principal findings through rigorous experimental analysis: 
\begin{enumerate*}[leftmargin=*,itemsep=0pt,topsep=0pt,parsep=0pt] 

\item \textbf{zero-shot text retrieval setting}: 
Code-specialized LLMs and the vision-language model demonstrated superior performance compared to other LLMs. 
This highlights the advantage of multimodal understanding and code-centric pretraining in text retrieval scenarios. 

\item \textbf{zero-shot code retrieval setting}: 
Vision-language models and code-specialized LLMs demonstrate superior zero-shot performance compared to other LLMs, even surpassing BM25. 

\item \textbf{supervised training setting}: 
After supervised training, the vision-language model and code-specialized LLMs maintain comparable performance to base models. 
Other specialized LLMs exhibited significant degradation in retrieval tasks. 
\end{enumerate*} 

\textbf{In summary}: 
The long-reasoning LLM and mathematics-specialized LLMs consistently underperformed across all three settings. 
This implies that models optimized for logical reasoning tasks may lack effectiveness in global semantic retrieval, which prioritizes contextual understanding over stepwise deduction. 
Moreover, the robust performance of the vision-language model and code-specialized LLMs suggests promising directions for the unified retrieval task leveraging cross-domain and cross-modal fusion.

\section{Related Work} 
\heading{Passage Retrieval} Passage retrieval focuses on finding relevant information from extensive passage repositories based on the search queries using retrieval models. 
Conventional retrieval approaches such as TF-IDF and the probabilistic BM25 model \cite{robertson2009probabilistic} primarily rely on lexical term matching and inverse document frequency calculations for relevance measurement. 
While these algorithms demonstrate notable efficiency in retrieval operations, they are fundamentally constrained by lexical mismatch issues \cite{guo2022semantic}. 
Dense Retrieval (DR) becomes another popular retrieval model, encoding queries and documents into dense vector representations and measuring relevance through similarity computations. 
This section will briefly introduce different dense retrieval models: pre-trained language models-based dense retriever (PLM-based DR) and large language models-based dense retriever (LLM-based DR). 
\subsection{PLM-based DR}
To further enhance dense retrieval methods based on pre-trained language models, advancements have been made in two key areas: 
\begin{enumerate*}[leftmargin=*,itemsep=0pt,topsep=0pt,parsep=0pt] 
    \item customized pre-training for dense retrieval \cite{xiao2022retromae,chang2020pre,ma2021prop}: 
    For example, RetroMAE \cite{xiao2022retromae} employed masked auto-encoder (MAE) frameworks tailored for IR. These methods introduce a shallow decoder to reconstruct the original input from masked tokens and sentence embeddings, enhancing the model’s ability to generate meaningful representations for retrieval purposes. 
    Other works, such as those by \cite{chang2020pre}, propose paragraph-level pre-training tasks like the Inverse Cloze Task (ICT) to better capture contextual relationships between queries and documents.
    
    \item hard negative mining: 
    For example,
    RocketQA \cite{qu2020rocketqa} enhances retrieval performance through cross-batch negatives and denoised hard negatives. 
\end{enumerate*}

\subsection{LLM-based DR}
The emergence of Large Language Models (LLMs) like Qwen \cite{bai2023qwen}, and GPT-4 \cite{gpt-4} has demonstrated exceptional performance in various domains, particularly in natural language generation. 
Recent studies \cite{ma2024fine,echo} have explored incorporating decoder-only LLM architectures into dense retrieval systems, encoding queries and documents into embeddings for similarity comparison.
For instance, RepLLaMA \cite{ma2024fine} replaces PLMs with LLMs for encoding and shows that even with simple fine-tuning strategies, LLM-based retrievers can surpass PLM-based retrievers that employ more complex training methods. 
LLM2VEC \cite{behnamghader2024llm2vec} modifies the decoder-only architecture to include bidirectional attention mechanisms, further improving retrieval effectiveness. 
Llama2Vec \cite{li2024llama2vec} introduces pre-training tasks such as Embedding-based Auto-Encoding and Embedding-based Auto-Regression to better adapt LLMs for dense retrieval applications. Echo \cite{springer2024repetition} tackles architectural limitations by repeating inputs in the context and extracting embeddings from subsequent occurrences.
LLM-QL \cite{zhang2025unleashing} proposed query likelihood modeling training before conservative learning to enhance retrieval performance. 

\section{LLM-based Retrieval}

Dense Retrieval (DR)  is a retrieval framework that learns dense vector representations of queries and passages through dual-tower encoders. Given a query $q$ and a passage $p$, the model computes their embeddings as: 

\begin{equation}
\mathbf{h}_q = E_Q(q) \in \mathbb{R}^d, \quad \mathbf{h}_p = E_P(p) \in \mathbb{R}^d
\end{equation}

where $E_Q(\cdot)$ and $E_P(\cdot)$ denote the query and passage encoders, respectively, and $d$ is the embedding dimension.

\heading{Training Strategy: Contrastive Learning}
The model is trained to maximize the similarity between queries and relevant passages while minimizing similarity with non-relevant ones. For each query-positive passage pair $(q, p^+)$, we construct negative examples $\{p_i^-\}_{i=1}^N$ containing both hard negatives and in-batch negatives. 
The training is then optimized according to contrastive learning: 
\begin{equation}
\mathcal{L} = -\log \frac{e^{s(q, p^+)}}{e^{s(q, p^+)} + \sum_{i=1}^N e^{s(q, p_i^-)}}
\end{equation}

where $s(q, p) = <\mathbf{h}_q, \mathbf{h}_p>$ is the similarity score between the query and document measured by dot product or cosine similarity.







\section{Experimental Setup} 

\subsection{LLMs}
\heading{Qwen2.5}
Qwen2.5 \cite{yang2024qwen2} maintains the Transformer-based decoder architecture \cite{vaswani2017attention} of Qwen2 \cite{anyang2024qwen2}. Compared to Qwen2, Qwen2.5 has a larger and higher-quality pre-training dataset, expanding from 7 trillion tokens to 18 trillion tokens. 
The pre-training data spans various domains, including e-commerce, social media, and entertainment, as well as mathematical data from Qwen2.5 Math \cite{yang2024qwen2math} and coding data from Qwen2.5 Coder \cite{hui2024qwen2coder}. The Qwen2.5-Instruct model is derived from Qwen2.5 through post-training involving supervised fine-tuning and reinforcement learning.

\heading{Qwen2.5 Math}
Qwen2.5 Math \cite{yang2024qwen2math} is obtained by further pre-training the Qwen2.5 model on high-quality mathematical data, specifically the Qwen Math Corpus v2, which contains over 1 trillion tokens. After completing this pre-training, Qwen2.5-Math-Instruct is fine-tuned using Chain-of-Thought (CoT) and Tool-Integrated Reasoning (TIR) methods on the Qwen2.5 Math base model.

\heading{Qwen2.5 Coder}
Qwen2.5 Coder \cite{hui2024qwen2coder} is trained on a large-scale, coding-specific pre-training dataset comprising over 5.5 trillion tokens, sourced from a wide range of public code repositories such as GitHub and extensive web-crawled data containing code-related texts. The dataset includes five types: Source Code Data, Text-Code Grounding Data, Synthetic Data, Math Data, and Text Data. Balancing code, math, and text data is crucial for building a foundational model; it was found that a 7:2:1 ratio outperformed other configurations. Qwen2.5-Code-Instruct is fine-tuned using Supervised Fine-Tuning (SFT) and Direct Preference Optimization (DPO) on the Qwen2.5 Coder base model.


\heading{DeepSeek-R1-Distill-Qwen}
DeepSeek-R1-Distill-Qwen \cite{deepseek-r1}, obtained by directly distilling DeepSeek-R1 into the Qwen2.5 \cite{yang2024qwen2} model, outperforms the approach of applying reinforcement learning to Qwen2.5.  

\heading{Qwen2.5 VL} 
Qwen2.5-VL \cite{qwen2.5vl} introduces significant vision-language advancements over its predecessor, including enhanced omnidocument parsing for multilingual, multi-scene texts and specialized formats (tables, formulas, etc.), improved object grounding accuracy with spatial reasoning in JSON/coordinate formats, and ultra-long video understanding via dynamic resolution/frame-rate training for temporal event extraction. 
Specifically, Qwen2.5-VL contains a large language model that is initialized with pre-trained weights from the Qwen2.5 LLM, a vision encoder that is a redesigned Vision Transformer (ViT) architecture, and a two-layer multi-layer perceptron(MLP) that projects vision features into a dimension that aligns with the text embeddings used in the LLM. 

\vspace{-3mm}
\subsection{Datasets} 


\heading{CoIR Benchamrk} 
To verify the domain capability on retrieval performance, we evaluate code retrieval datasets on the CoIR benchmark \cite{CoIR}: 
1) Text-to-Code Retrieval (T2C): 
a. APPS \cite{apps}, which is a diverse collection of problems from platforms; 
b. CosQA \cite{cosqa}, which contains labeled pairs of textual queries and Python functions; 
c. Synthetic-Text2Sql, which is the largest and most diverse synthetic dataset. 
2) Code-to-Text Retrieval (C2C): 
a. CodeTrans \cite{Codetransocean}, which contains code written within the same programming language, i.e., CodeTrans-Contest, and different programming languages, i.e., CodeTrans-DL;
b. CodeSearchNet-CCR, which is the modified dataset from the original CodeSearchNet. 
3) Code-to-Text Retrieval (C2T):
a. CodeSearchNet \cite{husain2019codesearchnet}, which consists of numerous code functions accompanied by code comments.
3) Hybrid Code Retrieval (Hybrid): 
a. Stackoverflow-QA\footnote{\url{https://www.kaggle.com/datasets/stackoverflow/stacksample/data}}, which is derived from the original StackOverflow dataset by pairing questions with their highest upvoted answers; 
b. Codefeedback \cite{Opencodeinterpreter}, which is a synthesized code instruction dataset generated by LLMs, contains Codefeedback-MT and CodeFeedBack-ST.

\heading{BEIR Benchmark} We consider low-resource datasets from the BEIR  \cite{beir}, so we employ six datasets, consisting of question-answering task: FiQA \cite{fiqa} and Quora; citation-prediction task: SCIDOCS \cite{scidocs}; bio-medical IR task: TREC-COVID \cite{trec-covid} and NFcorpus \cite{nfcorpus}; argument retrieval task: ArguAna \cite{arguana}. 

\heading{MS MARCO}
In our experiments, we utilize the MS MARCO passage retrieval dataset \cite{msmarco2016} for both training and evaluation purposes. 
We train different Qwen models on the MS MARCO training set, which includes about 500K queries. 
For evaluation, we employ three datasets: MS MARCO-dev, TREC-DL19 and TREC DL20 \cite{trecl19}. 
The MS MARCO-dev set consists of 6,980 queries, each associated with an average of 1.1 relevant passages. 


\vspace{-3mm}
\subsection{Zero-shot Details} 
For zero-shot retrieval performance evaluation, the design of prompts significantly impacts model performance. 
We adopt the prompts from previous work \cite{Promptreps}, which have been proven effective in text retrieval tasks. The specific prompts are formulated as follows:
1. \textbf{Prompt for document:}
``Passage: ``[text]''. Use one word to represent the passage in a retrieval task. Make sure your word is in lowercase. <A>The word is: ``''. 
2. \textbf{Prompt for query:}
Prompt for query representation:
``Query: ``[text]''. Use one word to represent the query in a retrieval task. Make sure your word is in lowercase. <A>The word is: ``''. ``<A>'' denotes the model-specific assistant special token. The hidden state of the last token `` as the query or document embedding. 
To ensure fair comparison across all zero-shot retrieval tasks, identical prompts are consistently used for all LLMs. 

\subsection{Training Details} 
All Qwen models are trained with Deepseed, which is an efficient deep-learning optimization library, and  Zero Redundancy Optimizer-3 (ZeRO-3), which is a family of memory optimization technologies for large-scale distributed deep learning.  
For hard negative mining, we adopt the methodology established in RepLLaMA \cite{ma2024fine}, implementing a hybrid approach combining BM25 lexical retrieval with CoCondenser \cite{gao2021unsupervised} dense retrieval. This dual-channel selection mechanism ensures comprehensive coverage of challenging negative samples from both sparse and dense representation spaces. 
Following \cite{ma2024fine}, the training protocol employs a single epoch with a global batch size of 32 (achieved through 4 gradient accumulation steps on 8 NVIDIA A800 GPUs with 80GB memory). 
The optimization process utilizes the AdamW optimizer with an initial learning rate of 1e-4, maintaining hardware utilization efficiency through ZeRO-3 parallelism. 
We use the parameter-efficient tuning method, i.e., LoRA with the  LoRA rank of 32. 
Moreover, we can extract the final-layer hidden state representation of the end-of-sequence (EOS) special token in LLMs as the dense representation for the query or document.
The prompts for the query and document are ``Query:'' and ``Passage:'', respectively.   

\section{Experiment Results}


\subsection{Zero-shot Text Retrieval Results}
\begin{table}[t]
  \centering
  \small
  \caption{Zero-shot NDCG@10 retrieval performance on text retrieval datasets. ``Base'', ``Instruct'', ``Coder'', ``Coder-I'', ``Math'', ``Math-I'', ``Distill'', and ``VL-I'' are ``Qwen2.5-7B'', ``Qwen2.5-7B-Instruct'',	``Qwen2.5-coder-7B'',	``Qwen2.5-coder-7B-Instruct'',	``Qwen2.5-math-7B'',	``Qwen2.5-math-7B-Instruct'',	``Deepseek-R1-Distill-Qwen-7B'',	and ``Qwen2.5-VL-7B-Instruct'', respectively. \textbf{Bold} means the best performance. }
    \begin{tabular}{llllllllll}
    \toprule
    Dataset & \multicolumn{1}{l}{BM25} & \multicolumn{1}{l}{Base} & \multicolumn{1}{l}{Instruct} & \multicolumn{1}{l}{Coder} & \multicolumn{1}{l}{Coder-I} & \multicolumn{1}{l}{Math} & \multicolumn{1}{l}{Math-I} & \multicolumn{1}{l}{Distill} & \multicolumn{1}{l}{VL-I} \\
    \midrule
    FiQA  & 0.2361  & 0.1239  & 0.1819  & 0.2192  & 0.2482  & 0.0731  & 0.1192  & 0.1237  & \textbf{0.2531}  \\
    SCIDOCS & 0.1490  & 0.1356  & 0.1365  & 0.1527  & \textbf{0.1784}  & 0.1302  & 0.0853  & 0.0976  & 0.1601  \\
    TREC-COVID & 0.5947  & 0.5622  & 0.5060  & 0.5566  & 0.5643  & 0.3869  & 0.4364  & 0.4181  & \textbf{0.6094}  \\
    NFcorpus & \textbf{0.3218}  & 0.2045  & 0.2237  & 0.1928  & 0.2793  & 0.1148  & 0.1673  & 0.1200  & 0.2447  \\
    Quora & 0.7886  & 0.7202  & 0.6936  & \textbf{0.8024}  & 0.7952  & 0.5211  & 0.6834  & 0.7199  & 0.7799  \\
    ArguAna & \textbf{0.3970}  & 0.3229  & 0.1882  & 0.3304  & 0.2772  & 0.2926  & 0.1696  & 0.2195  & 0.2507  \\
    \midrule
    Avg   & \textbf{0.4145}  & 0.3449  & 0.3217  & 0.3757  & 0.3904  & 0.2531  & 0.2769  & 0.2831  & 0.3830  \\
    \bottomrule
    \end{tabular}%
  \label{tab:zero-shot-text-retrieval}%
\end{table}%
We evaluate the zero-shot NDCG@10 retrieval performance of various specialized LLMs against the BM25 baseline across six datasets from the BEIR benchmark, as shown in Table \ref{tab:zero-shot-text-retrieval}. Key observations are summarized as follows: 

1) \textbf{Domain-specialized LLMs outperform general-purpose LLMs}.
Models with code-specific pertaining and multimodal tuning demonstrate superior retrieval capabilities. The coder-oriented variants (Coder: 0.3757, Coder-I: 0.3904) and vision-language model (VL-I: 0.3830) achieve average performance exceeding the base model (0.3449), with Qwen2.5-VL-7B-Instruct showing best performance on biomedical (trec-covid: 0.6094) and financial (FiQA: 0.2531) queries. 
This suggests that specific domain adaptation for LLMs can enhance zero-shot retrieval effectiveness. 

2) \textbf{Instruction tuning has variable impacts}.
The effect of instruction tuning varies significantly across model specializations. 
Instruction tuning of base models demonstrates a pronounced adverse impact on retrieval task performance. 
However, domain-adapted LLMs necessitate subsequent instruction tuning to attain improved comprehension of retrieval tasks. 
For example, while instruction tuning is beneficial for coder models (+3.9\% Coder→Coder-Instruct and 9.4\% Math→Math-Instruct), it degrades performance in base models (-6.7\% Base→Instruct). 

3) \textbf{Long reasoning reduces retrieval performance}.
While test-time scaling demonstrates substantial advantages in question answering and related tasks, it has been observed to exhibit markedly diminished performance in retrieval scenarios. 
For example, the retrieval performance of the long reasoning LLM decreased by 17.9\% compared to the base model in terms of NDCG@10. 
This discrepancy may stem from fundamental task objective misalignment: retrieval systems primarily require global semantic alignment, whereas question-answering and reasoning tasks demand granular comprehension of contextual details.

4) \textbf{Mathematical specialization hinders general retrieval}. 
Mathematical LLM demonstrates the weakest performance, underperforming even Deepseel-R1-Distill-Qwen-7B. 
This observation suggests that retrieval tasks may not necessitate sophisticated logical-mathematical reasoning capabilities.

\textbf{In summary}. 
Coding specialized pertaining and multimodal capability significantly impacts zero-shot retrieval effectiveness. 
Long reasoning and mathematical conflicts in global semantic matching result in worse retrieval performance. 
These findings suggest coding specialized pertaining and multimodal capabilities as promising directions for generalizable text retrieval tasks.

\subsection{Zero-shot Code Retrieval} 
\begin{table}[t]
  \centering
  \small
  \caption{Zero-shot NDCG@10 retrieval performance  on code retrieval datasets. \textbf{Bold}, ``Base'', ``Instruct'', ``Coder'', ``Coder-I'', ``Math'', ``Math-I'', ``Distill'', and ``VL-I'' are defined in Table \ref{tab:zero-shot-text-retrieval}. ``S-T'', ``CF-'', ``CT-''  ``CS'', ``CT-C'' and ``SK-'' are the ``Synthetic-Text2Sql'', ``CodeFeedBack-'', ``CodeTrans-'', ``CodeSearchNet'', ``CodeTrans-Contest'' and ``Stackoverflow-'', respectively. }
  \setlength\tabcolsep{1pt}
  \renewcommand{\arraystretch}{0.95}
    \begin{tabular}{cllllllllll}

    \toprule
    \multicolumn{1}{l}{Task} & Dataset & \multicolumn{1}{l}{BM25}  & \multicolumn{1}{l}{Base} & \multicolumn{1}{l}{Instruct} & \multicolumn{1}{l}{Coder} & \multicolumn{1}{l}{Coder-I} & \multicolumn{1}{l}{Math} & \multicolumn{1}{l}{Math-I} & \multicolumn{1}{l}{Distill} & \multicolumn{1}{l}{VL-I} \\ \\
    \midrule
     \multirow{3}[2]{*}{T2C} & \multicolumn{1}{l}{APPS} & 0.0095  & 0.0162  & 0.0774  & 0.0732  & 0.1611  & 0.0325  & 0.0773  & 0.0449  & \textbf{0.1948}  \\
          & \multicolumn{1}{l}{CosQA} & 0.1396  & 0.0759  & 0.2128  & 0.2193  & 0.2677  & 0.1213  & 0.1750  & 0.1646  & \textbf{0.2519}  \\
          & \multicolumn{1}{l}{S-T} & 0.1692  & 0.2880  & 0.0863  & \textbf{0.5135}  & 0.4840  & 0.3380  & 0.2646  & 0.2268  & 0.2874  \\
    \midrule
    C2T   & \multicolumn{1}{l}{CS} & 0.2675  &   0.2632    & 0.0891  & \textbf{0.3737}  & 0.3699  & 0.1584  & 0.2260  & 0.2619  & 0.2067  \\
    \midrule
    \multirow{3}[2]{*}{C2C} & \multicolumn{1}{l}{CT-C} & 0.5013  & 0.2320  & 0.2756  & 0.3163  & 0.3453  & 0.1753  & 0.3105  & 0.1456  & \textbf{0.7813}  \\
          & \multicolumn{1}{l}{CT-DL} & 0.0869  & 0.1599  & 0.1706  & 0.1602  & 0.1749  & 0.1007  & 0.1471  & 0.1883  & \textbf{0.2942}  \\
          & \multicolumn{1}{l}{CS-CCR} & \textbf{0.3469}  &  0.2123     & 0.1520  & 0.2760  & 0.3005  & 0.1544  & 0.2001  & 0.1720  & 0.3360  \\
    \midrule
    \multirow{3}[2]{*}{Hybird} & \multicolumn{1}{l}{SK-QA} & 0.5680  & 0.4000  & 0.3069  & 0.4790  & 0.4926  & 0.2109  & 0.2798  & 0.2535  & \textbf{0.6169}  \\
          & \multicolumn{1}{l}{CF-MT} & 0.3473  & 0.0758  & 0.0701  & 0.1126  & 0.0938  & 0.0519  & 0.0698  & 0.0481  & \textbf{0.5070}  \\
          & \multicolumn{1}{l}{CF-ST} & 0.5432  & 0.2378  & 0.2962  & 0.4699  & 0.4468  & 0.2284  & 0.3096  & 0.2612  & \textbf{0.5666}  \\
    \midrule
    \multicolumn{2}{c}{Avg} & 0.2979  & 0.1961  & 0.1737  & 0.2994  & 0.3137  & 0.1572  & 0.2060  & 0.1767  & \textbf{0.4043}  \\
    \bottomrule
    \end{tabular}%
  \label{tab:code_retrieval}%
\end{table}%

Table \ref{tab:code_retrieval} examines the zero-shot NDCG@10 performance of specialized LLMs for code retrieval on the CoIR benchmark. 
Key findings are summarized as follows: 

1) \textbf{Multimodal pretraining achieves best code retrieval performance}. 
The vision-language model Qwen2.5-VL-7B-Instruct achieves the best code retrieval performance and outperforms BM25 by 48\% in terms of average NDCG@10. 
This suggests that cross-modal training enhances code representation learning through improved structural understanding.

2) \textbf{Code-specialized LLMs achieve superior code retrieval performance}. 
Models pre-trained on code-centric objectives demonstrate significant advantages compared to the base LLM. 
The average performance of the instruction-tuned coder variant is improved by 66\% compared to the base LLM. 
This highlights the critical role of code-aligned pretraining for semantic code matching.

3) \textbf{Instruction tuning amplifies code retrieval capabilities}.  
Instruction tuning consistently improves code retrieval performance across different domain-specific LLMs, e.g., coder, and math LLMs. 
This demonstrates the generalizability of instruction tuning for zero-shot code retrieval tasks.

4) \textbf{Mathematical pretraining shows worst code retrieval Performance}. 
Similar to text retrieval performance, the math-adaption LLM has the worst performance, which further indicates that limited cross-domain transfer between mathematical reasoning and semantic matching. 

5) \textbf{Long reasoning compromises code retrieval performance}.  
The performance of the long reasoning LLM is decreased by 10.3\% compared to the base LLM in terms of NDCG@10, suggesting reasoning techniques may disproportionately affect code retrieval, which is similar to the text retrieval task.

\textbf{In summary}. The vision-language model and code-specialized LLMs achieve superior zero-shot code retrieval performance compared to other LLMs, even compressing BM25. 
The experimental findings substantiate that cross-modal interactions and code structure comprehension exhibit alignment, thereby enabling global semantic matching. 
Similar to zero-shot text retrieval, the long reasoning LLM and mathematical LLMs show worse retrieval performance, further indicating that reasoning and mathematical logic are not alignment with global semantic matching. 
The observed performance decline in the long-reasoning LLM likely stems from missing intermediate reasoning steps, though untested due to computational constraints, warranting further investigation.


\subsection{Supervised Performance}
\begin{table}[t]
  \centering
  \small
  \caption{Different retrieval performance using different LLMs on MS MARCO and TREC DL 19/20. ``$^-$'' indicate significant  decrements over Qwen2.5-7B, using a two-sided
paired t-test ($p$ < 0.05). \textbf{Bold} indicates the best performance.}
    \begin{tabular}{llllllll}
    \toprule
    \multirow{2}[4]{*}{LLM} & \multicolumn{5}{c}{DEV}       & DL19 & DL20 \\
\cmidrule(r){2-6} \cmidrule(r){7-8}          & M@10 & N@10 & R@100 & R@200 & R@1000 & N@10 & N@10 \\
    \midrule
    Qwen2.5-7B & \textbf{0.423}  & 0.491  & 0.945 & \textbf{0.973}  & \textbf{0.995}  & 0.734  & 0.723  \\
    Qwen2.5-7B-instruct & \textbf{0.423}  & \textbf{0.492}  & \textbf{0.945} & \textbf{0.973} & 0.994  & \textbf{0.740}  & 0.729  \\
    Qwen2.5-coder-7B & 0.419  & 0.489  & 0.944 & 0.970 & 0.995  & 0.733  & \textbf{0.742}  \\
    Qwen2.5-coder-7B-instruct & 0.418  & 0.488  & 0.944 & 0.969$^-$  & 0.995  & 0.733  & 0.734  \\
    Qwen2.5-math-7b & 0.415$^-$  & 0.484$^-$  & 0.939$^-$ & 0.965$^-$  & 0.993$^-$  & 0.738  & 0.721  \\
    Qwen2.5-math-7b-instruct & 0.413$^-$  & 0.480$^-$  & 0.934$^-$ & 0.964$^-$ & 0.992$^-$  & 0.727  & 0.715  \\
    Deepseek-R1-distill-Qwen-7B & 0.413$^-$  & 0.481$^-$  & 0.935$^-$  & 0.966$^-$ & 0.992$^-$  & 0.737  & 0.703  \\
    Qwen2.5-VL-7B-instruct & 0.418$^-$  & 0.488  & 0.943 & 0.969$^-$ & 0.994  & 0.727  & 0.739  \\
    \bottomrule
    \end{tabular}%
  \label{tab:supervised_performance}%
\end{table}%
Table \ref{tab:supervised_performance} shows the supervised retrieval performance upon different LLMs. 
We can  observe that: 

\textbf{Capability specialization impacts}.
Retrievers upon different specialized LLMs after contrastive learning reveal distinct retrieval performance. 
Coding-oriented variants and the vision-language model maintain comparable performance to the general base LLM on the DEV set on most metrics.
Qwen2.5-coder-7B retriever achieves the best performance on TREC DL20 and has a similar performance on TREC DL19 with Qwen2.5-7B retriever. 
However, mathematical and reasoning specialization causes consistent degradation across all DEV metrics (e.g., -1.8\% NDCG@10 for Qwen2.5-math-7b). 
Moreover, Qwen2.5-math-7B-instruct and Deepseek-R1-distill-Qwen-7B retrievers also achieve the worst performance on TRECL DL19 and TREC DL20, respectively. 
This suggests that mathematical abstraction may conflict with global semantic matching requirements in text retrieval tasks. 
The reasoning LLMs, i.e., Deepseek-R1-distill-Qwen-7B, which are effective in generation tasks such as question answering, have significantly reduced performance in retrieval tasks after supervised training. 
The possible reason is that retrieval tasks do not require complex reasoning but rely on the representation of global semantics.

\vspace{-2mm}
\section{Conclusion}
We present a comprehensive investigation into the relationship between LLMs' specialty and their retrieval performance across supervised and zero-shot settings. 
We conduct extensive experiments using eight Qwen2.5-7B LLMs on zero-shot text/code settings and supervised settings.  
We observe three key findings: 
(1) Mathematical specialization and long reasoning LLMs cause consistent degradation in all three settings, indicating conflicts between mathematical reasoning and semantic matching; 
(2) Code-specialized and vision-language models demonstrate superior zero-shot performance compared to other LLMs and maintain comparable performance to base LLMs in supervised settings. 
The robust performance of the vision-language model and coding-specialized LLMs suggests promising directions for the unified retrieval task leveraging cross-domain and cross-modal fusion. 
The significant performance degradation of long-reasoning LLMs across three settings may potentially stem from the absence of the model to output its intermediate reasoning (thinking) process. 
However, due to the computational costs associated with generating extensive outputs from LLMs, this aspect was not tested in the current study and merits further investigation in future work.

\section{Acknowledgment}
This work was funded by the National Natural Science Foundation of China (NSFC) under Grants No. 62302486, the Innovation Project of ICT CAS under Grants No. E361140, the CAS Special Research Assistant Funding Project, the project under Grants No. JCKY2022130C039, the Strategic Priority Research Program of the CAS under Grants No. XDB0680102, and the NSFC Grant No. 62441229.

%
%
%
%
\bibliographystyle{splncs04}
\bibliography{reference}
\end{document}